# P2P SIMULATOR FOR QUERIES ROUTING USING DATA MINING

Anis ISMAIL[1], Aziz BARBAR[1] and Ziad ISMAIL[2]

[1] American University of Science & Technology
Alfred Naccash Avenue - Ashrafieh
Beirut, Lebanon
`anismaiil@ul.edu.lb`
`abarbar@aust.edu.lb`
[2] TELECOM ParisTech
Paris, France
`ismail.ziad@enst.fr`

## ABSTRACT

*Data mining is used to extract hidden information from large databases. In Peer-to-Peer context, a challenging problem is how to find the appropriate Peer to deal with a given query without overly consuming bandwidth. Different methods proposed routing strategies of queries taking into account the P2P network at hand. An unstructured P2P system based on an organization of Peers around Super-Peers that are connected to Super-Super-Peer according to their semantic domains is considered. This paper integrates Decision Trees in P2P architectures for predicting Query-Suitable Super-Peers representing a community of Peers, where one among them is able to answer the given query. In fact, by analyzing the queries' log file, a predictive model that avoids flooding queries in the P2P networks constructed by predicting the appropriate Super-Peer, and hence the Peer to answer the query. The proposed architecture is based on a Decision Tree (Base-Knowledge - BK). The efficiency of these architectures is discussed considering architecture without knowledge (Baseline) using only the flooding queries method to answer queries. The advantage of this knowledge based model is the robustness in Queries routing mechanism and scalability in P2P Network.*

## 1. INTRODUCTION

With the rapid increase in the number of computers connected to the Internet and the emergence of a range of mobile computational devices which might soon be equipped with mobile IP technology, the Internet is converging to a more dynamic, huge, extremely heterogeneous network which provides basic services such as routing and name lookup. This platform is already being used to support huge, fully distributed Peer-to-Peer overlay networks containing millions of nodes, typically for the purpose of information dissemination and file sharing [1][2][3]. Such fully distributed systems generate immense amounts of data. Analyzing this data can be interesting for both scientific and business purposes. Among other applications, this environment is a natural target for distributed data mining [4].

Data mining has recently become very popular due to the emergence of vast quantities of data. In this paper, potential pitfalls and practical issues about data mining P2P network are discussed. The motivations behind P2P data mining include the optimal usage of available computational resources, privacy and dependability by eliminating critical points of service. We will adopt the harshest possible constraints on the distribution of data and the elements of the network, and demonstrate techniques which can still provide useful information about the distributed data effectively and dependably.

Two constraints will be adopted. The first is that all Peers are allowed to hold little data, and are called "Expertise". The second doesn't have practically a limited number of nodes; basically, the only requirement is that each Peer could communicate directly with its Super-Peer which is the parent of the domain. Furthermore, we will concentrate on two other very important aspects: data privacy and the dynamic nature of the underlying network (Peers can leave the overlay network and new Peers can join it).

To achieve our goal, we suggest a new system that uses Decision Trees to extract Super-Peers that contain Peers relevant to a given query. This system is an unstructured P2P system based on an organization of Peers around Super-Peers that is connected to Super-Super-Peer according to their semantic domains, and which also uses Decision Trees to extract Super-Peers that contains Peers with relevant data that respect a given query. The advantage of this model is the robustness in Queries Answering and scalability issues in P2P Network.

This paper is divided into 5 sections; section 1 offers a literature review of the Peer-to-Peer system; section 2 describes the system's architecture; section 3 represents our experiments protocol; section 4 evaluates our new system by running a series of experiments; and finally, section 5 provides the conclusion and proposes some future works.

## 2. RELATED WORK

Data mining (DM) deals with the problem of data analysis in environments with distributed data, computing nodes, and users. This area has witnessed considerable amount of research during the last decade. P2P data mining has recently emerged as an area of Distributed Data Mining (DDM) research, specifically focusing on algorithms which are asynchronous, communication-efficient and scalable.

Peer-to-Peer (P2P) systems are distributed systems without centralized control in which each node shares and exchanges data across a network. The salient features of P2P systems make up a long list: low-cost sharing of data, redundant storage, self-organization, autonomy, load-balancing and fault tolerance. A new category of P2P systems, called schema-based, in which each Peer is a database management system in itself and exposes its own schema, have recently drawn considerable attention. These systems can support distributed heterogeneous data stores, and combine approaches from P2P research and from the database and semantic web research areas.

However, such systems that broadcast all queries to all Peers suffer from limited efficiency and scalability. Intelligent routing strategies are essential in such settings, so that queries get only routed to a semantically chosen subset of Peers who are able to answer parts of or the whole queries. The problem of efficient query routing has been studied by many researchers since its solution affects the overall search mechanism of the P2P network. Moreover, in order to gain fast retrieval of data without large bandwidth consumption, there is a need for efficient ways of queries processing by a selected set of Peers.

The work described in this paper relates to two main bodies of research: Data mining and Query routing in P2P context.

## 2.1. Data mining in Peer-to-Peer Network

Knowledge discovery and data mining from P2P network is a relatively new field with little related literature. Some researchers have developed several different approaches for computing basic operations (e.g. average, sum, max, execution time) on P2P networks. For instance, Raahemi et al. [5] presented a new approach using data-mining technique, in particular Decision Tree, to classify Peer-to-Peer (P2P) traffic in IP networks by capturing Internet traffic at a main gateway router, perform data pre-processing, select the most significant attributes, and prepare a training-data set to which the decision-tree algorithm to be applied. They built several models using a combination of various attribute sets for different ratios of P2P to non-P2P traffic in the training data. They observed that the accuracy of the model increased significantly when they included the attributes "Src IP addr" and "Dst IP addr" when building the model.

A different complexity involves the usage of meta-data, which was shown to be particularly useful for finding similarity between performing artists [6]. The content on file sharing networks is mostly ripped by individual users for consumption by other users. User based interactions are a desirable property in Information Retrievals (IR) data-sets; however, when it comes to meta-data, it's the main source for ambiguities and noise. Be it a movie, a song, or any other file type, typically there would be several similar duplications available on the network. The files may be digitally identical, thus having the same hash signature, yet bearing different file names and meta-data tags. Duplication in meta-data tags is typically caused by spelling mistakes, missing data, and different variations of the correct values. In the Gnutella network, for example, only 7-10% of the queries are successful in returning useful content [7]. A common hash signature can facilitate similar files grouping; nonetheless, it does not solve the problem of copies that are not digitally identical. The problem of meta-data ambiguities in P2P data-set is addressed by Koenigstein et al. [8].

Roussopoulos et al [9] present a heuristic Decision Tree that designers can use to judge how suitable a P2P solution might be for a particular problem. It is based on characteristics of a wide range of P2P systems gleaned from the literature, both proposed and deployed. These include budget, resource relevance, trust, rate of system change, and criticality. Bhaduri et al [10] propose a P2P Decision Tree induction algorithm in which every Peer learns and maintains the correct Decision Tree as compared to a centralized scenario. This algorithm is completely decentralized, asynchronous, and adapts smoothly to changes in the data and the network. Bhaduri et al [10] offer a scalable and robust distributed algorithm for Decision Tree induction in large Peer-to-Peer (P2P) environments. Computing a Decision Tree in such large distributed systems using standard centralized algorithms can be very communication-expensive and impractical because of the synchronization requirements. The problem becomes even more challenging in the distributed stream monitoring scenario where the Decision Tree needs to be updated in response to changes in the data distribution. It presents an alternate solution that works in a completely asynchronous manner in distributed environments, and offers low communication overhead; a necessity for scalability.

Classification based on Decision Trees is one of the important problems in data mining and has applications in many fields. Bar-Or et al. [11] presents an algorithm that sharply reduces the communication overhead by sending just a fraction of the statistical data. They execute ID3 in a hierarchical network by centralizing, for every node of the tree and at each level, only statistics regarding the most promising attributes.

Content location is a challenging problem in decentralized Peer-to-Peer systems. And query-flooding algorithm in Gnutella system suffers from poor scalability and considerable network overhead. Currently, based on the Small-world pattern in the P2P system, a piggyback algorithm called interest-based shortcuts gets a relatively better performance. However, Xi Tonget al. [12]

believed that the said algorithm could be improved, and even become more efficient; hence, a cluster-based algorithm is put forward. The main concern of their algorithm is to narrow the search scope in content location. Resource shortcuts are grouped into clusters according to their contents, and resource queries are only searched in related shortcut clusters, so that the search efficiency is guaranteed and the network bandwidth is saved. In their experiment, cluster-based algorithm uses only 40% shortcuts roughly, as compared with the former algorithm, and the same success rate is achieved. At the end, they refer to the relationship between cluster-based algorithm and semantic overlay networks, which is a potential kind of overlay in the future.

Data mining over multiple data sources has emerged as an important practical problem with applications in different areas such as data streams, data-warehouses, and bioinformatics. Although the data sources are willing to run data mining algorithms in these cases, they do not want to reveal any extra information about their data to other sources due to legal or competition concerns. One possible solution to this problem is to use cryptographic methods. However, the computation and communication complexity of such solutions render them impractical when a large number of data sources are involved. F. Emekci et al. [13] consider a scenario where multiple data sources are willing to run data mining algorithms over the union of their data as long as each data source is guaranteed that its information, which does not pertain to another data source, will not be revealed. They focus on the classification problem in particular, and present an efficient algorithm for building a Decision Tree over an arbitrary number of distributed sources in a privacy preserving manner using the ID3 algorithm.

Koenigstein et al. [6] explore the relations between P2P and Billboard charts, showing a strong correlation between P2P queries and both Billboard Hot 100 and Digital Songs charts. They discussed how P2P queries reach their peak at the same time as a song reaches it highest Billboard ranking; thus, showing that P2P downloads and music sales are closely tied together, with little to no time gap. Yet, the P2P information is available a week before the Billboard charts are released. They suggest several novel prediction models of a song's success in the Billboard based on P2P queries and P2P popularity chart ranking. They manage to predict the success of a song in the Billboard Hot 100 with over 86% precision, and in Billboard Digital Songs with over 89% accuracy.

Medview [14] was designed earlier to support the learning process in oral medicine and oral pathology. The purpose of Medview was to provide a computerized teaching aid in these two domains. In this regard, a clinical database was created from the referrals; and, it has a large variation of clinical cases displayed by images, and test based information. The students reach the database through the media. They can practice and learn at any convenient time. Medview contains search tools to explore the database; accordingly, the students can study single cases or analyze various clinical parameters. MEduWeb is a web-based educational tool that allows students to search in the database and generate exercises with pictures of real patients [14]. MEduWebII was intended to enhance and improve mEduWeb program. It uses the MedView database that contains several thousand patient examinations; whereas, Khanet al. use Data mining technique (Decision trees) on this data base [15]. This work explores the possibilities of using Data mining technique (Decision trees) on the P2P database, and has performed a series of experiments within this context.

## 2.2. Query routing in P2P networks

Efficient query routing in P2P systems has already been discussed in the literature [16]. Semantic query routing techniques are required to improve effectiveness and scalability of search processes for resource sharing in P2P systems. The unstructured P2P systems typically employ flooding and random walk to locate data, which results in much network traffic. To improve their performances, classification based on Decision Trees that allows Peers to select the theme relevant to the queries is used.

The major problem in query routing in a P2P network is how the query is routed to a number of relevant Peers instead of being broadcasted to the whole network. This problem has been studied in recent works. Nejdl et al. [17] presented a routing strategy based on routing indices. Obviously, indices only help if they can exploit and express regularities present in the Peer and data distribution. A more advanced technique is presented by Löser et al. [18], [19] who introduced the notion of Semantic Overlay Clusters (SOCs) that define Peer clusters according to the Meta-data description of Peers and their contents. The former approach needs to accommodate index updates, whereas it needs an accurate definition of rules for Peers joining into a SOC. Both of these approaches use a Super-Peer topology. A Super-Peer is a node of the network that acts as a server to a subset of clients. This topology takes advantage of the heterogeneity of Peers; it is scalable as new Peers join and seems to be most suitable for schema-based P2P networks since it can support heterogeneous schema-based systems with different Meta-data schemas and ontologies.

In The Proceedings of the 4$^{th}$ International Conference on ontologies, Databases and Applications of Semantics [20], the INGA algorithm is presented. INGA extends the ideas of REMINDIN [21], where each Peer plays the role of a person in a social network. To determine the most appropriate Peers, each Peer maintains, in a lazy manner, a personal semantic shortcut index by analyzing the queries that are initiated by users of the P2P network and that happen to pass through the Peer. The main limitation of this routing approach is the unavoidable flooding of the network with messages when a new Peer (that has not yet stored any shortcuts) enters the network, or when Peers (in lower layers) contain limited information about queries that have already been answered in the past. The SQPeer routing strategy [22] uses intentional active schemas (RVL Views) for determining relevant Peer bases through the fragmentation of query patterns. However, since each view (active-schema) corresponds to a Peer advertisement, this view should be broadcasted to the whole P2P network.

## 3. SYSTEM ARCHITECTURE

### 3.1. Topology

The unstructured P2P system is based on an organization of Peers around Super-Peers according to their semantic domains (Figure 1). This proposal is based on the use of a distributed data structure, called expertise (p.r, r.m, m.i, h.i, ……), maintained by the Super-Peers, and describe data at the neighbouring Peers [4] [23] [24].

A Source (Peer) may send the query message to their Super-Peers which precisely pinpoint the pertinent Peer that belongs to this Peer, and to other Super-Peers that also contain pertinent Peer after processing the query. A challenging problem in a schema-based Peer-to-Peer (P2P) system is how to locate Peers that are relevant with respect to a given query with minimum query processing and answering time.

The proposed system (Figure 2) is an unstructured P2P system where Super-Peers are connected to a Super-Super-Peer that is the engine that specifies the Super-Peers that have the Peers which may have relevant data to answer to queries. This architecture combines centralized and unstructured approaches taking the advantages of centralized research and autonomy, and the distribution of loads and robustness for a distributed search. The Super-Peer architecture allows for the heterogeneity of Peers by assigning more responsibility to Peers who are able to assume them. Therefore, some Peer, called Super-Super-Peer (SSP), have an additional computing power and greater bandwidth, enabling them to perform administrative tasks. The SSP is responsible for managing all of Super-Peers; thus, reducing efforts of compilation of Queries, at the same time, preventing the spread of queries in the network. In each Theme, there is a Super-Peer which is connected to a Super-Super-Peer which has a global index that uses a Decision Tree to identify the Super-Peers that are most relevant to provide good results queries.

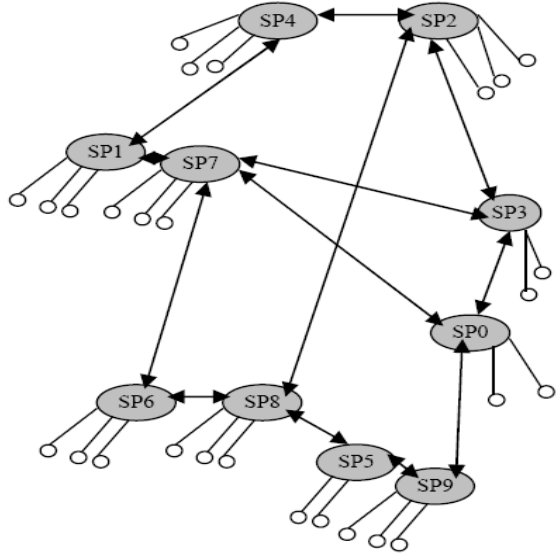

Figure 1. Semantic network of Super-Peer organized by themes (Baseline)

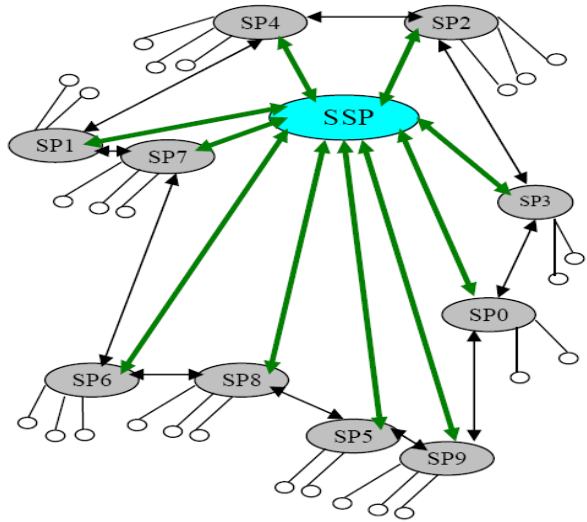

Figure 2. Semantic network of Super-Peer organized by themes with Super-Super-Peer (Base-Knowledge-BK)

### 3.2. SSP ARCHITECTURE

In this section, the logical architecture of a Super-Peer containing "knowledge", also known as SSP (Super-Super-Peer) is presented. This knowledge is represented by a Decision Tree. Decision trees are often used for classification and prediction. It is a simple and powerful knowledge representation.

An implementation of an existing algorithm in the WEKA platform is used to build the Decision Tree from the log file (log). The last component can predict, based on the Decision Tree, all Super-Peers who are able to respond to a query. The query manager receives a query from a Super-Peer in the domain-group (SSP) and returns the result of the prediction (the relevant Super-Peers) to this Super-Peer. Each SSP contains the following components:

- Query Manager: In this component, the routing application is based on the Decision Tree, allowing the prediction of Super-Peers to which candidates will forward the request to be processed.

- Logfile: It is the file that contains the queries processed by a Super-Peer domain. It contains the components of the application and the Super-Peer who has responded to the request. A line is added to the file Logfile of the domain, when a Super-Peer responds to a query.

- Method of construction of knowledge: The method is the algorithm for constructing the Decision Tree by analyzing the queries handled by domain-group members, and is stored in the Logfile. In the proposed experiments, the algorithm J48 WEKA platform to induce the Decision Tree is used.

- Prediction: This module uses the Decision Tree to predict, for a given query Q, the Super-Peers that are relevant candidates to process the query Q. Contrary to the general case where the tree used to predict a single value of the class (Super-Peer), we infer here all likely values of each class with its own probability. This list of class values is the set of Super-Peers that are likely to process the request.

## 3.3. Query routing by example

Assuming that Peer P1 issues a query Q, the query routing algorithm proceeds as follows:

- First, the responsible Super-Peer for P1 is found; in this example it is Super-Peer (SP1).

- The responsible Super-Peer sends the query to the Super-Super-Peer to identify the relevant Super-Peers for this query.

- The Super-Super-Peer will send the query to all relevant Super-Peers. Each relevant Super-Peer treats query to find relevant Peers.

- Then the final set of relevant Peers and their corresponding Super-Peers are returned.

## 4. SIMULATOR ARCHITECTURE

A simulator should allow us to experiment with different set-ups and configuration scenarios. In theory, a simulator should help us to test and develop our solution off-line and potentially with greater assurance. Since simulators can reduce the development time for a P2P system, it is important to review the various simulator implementations which provide support for P2P systems. Several Peer-to-Peer simulators exist:

OverlayWeaver [25] is a Peer-to-Peer overlay construction toolkit written in Java which can be used for easy development and testing of new overlay protocols and applications. The toolkit contains a so-called Distributed Environment Emulator which invokes and hosts multiple instances of Java applications on a single computer. This allows the simulation of up to 4,000 nodes. Since simulations have to be run in real-time and there is no statistical output, its use as an overlay network simulator is very limited.

PlanetSim [26] is an object-oriented simulation framework for overlay networks and services written in Java. It has a well-structured and modular architecture and makes use of the Common API [27]. In addition to the overlay protocols Chord [28] and Symphony [29] there are several services like CAST and DHT available on application layer. PlanetSim offers only limited support to collect statistics and has a very simplified underlying network layer without considering bandwidth and latency costs. This makes it difficult to simulate heterogeneous access networks and terminal mobility. It is possible to visualize the overlay topology at the end of a simulation run, but there is no interactive GUI.

A more comprehensive survey of Peer-to-Peer network simulators can be found in Proceedings of The Seventh Annual Postgraduate Symposium [29], where the authors show that most available Peer-to-Peer network simulators have several major drawbacks, limiting them in use for research projects.

For our implementation and simulation, we used the Java programming language, and the SimJava package. SimJava [30] is a process based discrete event simulation package for Java. Based on a discrete event simulation kernel, SimJava includes facilities for representing simulation objects as animated icons on screen. A SimJava simulation is a collection of entities each running in its own thread. These entities are connected together by ports and can communicate with each other by sending and receiving event objects.

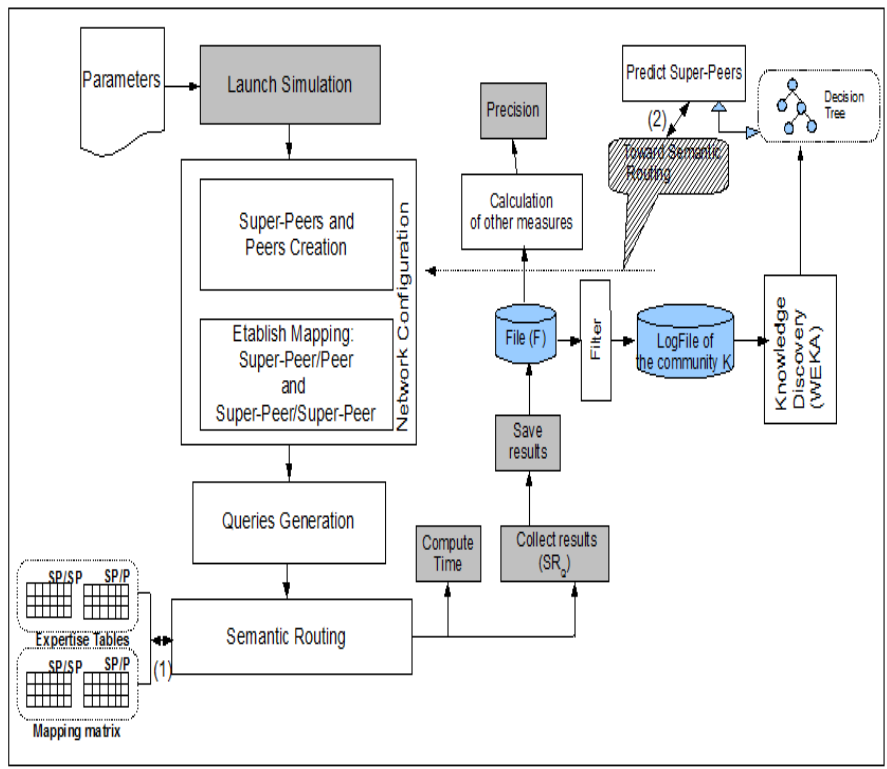

Figure 3. Simulator

Our simulator is based on a set of tools such as WEKA that is a data mining platform (See Figure 3).

We developed the necessary tools to interface the simulator with the various external components without any user intervention. This has led to the generation of a SON network with a level of Super-Peer (domain), which is juxtaposed to the Peer's level. A third level represents the knowledge using WEKA to predict the pertinent Super-Peer.

The simulation begins by sending queries; and, it is the characterizations of the third level to route queries to the relevant Super-Peers. The aggregation of all results returned by each Super-Peer that processed the query, contributes to the final result. The generation of applications is ensured by Peers. In fact, each Peer P can generate a query by selecting elements of expertise that become components of the query Q. We say that a Peer P is relevant to the query Q if the expertise of P contains at least a fraction of the components of Q. This is determined using the ability of a Peer P to resolve a query Q.

So each Peer generates a number N of queries that are derived from its expertise. After this phase, generation of query, Peers send their queries to their Super-Peers.

All queries exchanged within the network are stored in a file global LogFile. Thus, for a query Q, the file LogFile contains the following information: the identifier of the Peer (P), which submitted the application, its Super-Peer (SP), the query (Q) itself, and the Super-Peer which responded favourably to this request.

## 5. EXPERIMENTS PROTOCOL

We have used Weka [31] for our experiments. Weka is a collection of machine learning algorithms for data mining tasks, which was developed at the University of Waikato, New Zealand. It is written in Java and runs on almost any platform. The algorithms can either be applied directly to a dataset, or called from one's own Java code. Weka is also well-suited for developing new machine learning schemes. Weka is open source software issued under the GNU public license [31].

```
@relation P2P-BD
@attribute SuperPeer {SP0, SP1, SP2, SP3, SP4, SP5, SP6, SP7, SP8, SP9}
@attribute Query {Q10, Q11, Q12, Q13, ………………………….., Q308, Q309}
@attribute componentW1 {k.f, p.i, f.l, f.p, d.o, r.m, g.f, i.c, s.d, c.n, c.l, p.h, q.n, m.r, m.g, f.e, h.o, g.h,
j.l, o.h, j.m, b.e, h.j, g.m, k.m, h.n, p.n, d.e, p.r, e.s, l.n, j.h, f.k, h.s, h.i, m.i, m.l, l.e, i.o, k.l, r.k, o.f, g.c,
k.b, s.m, n.g, g.j, m.s, h.g, f.h, p.m, h.a, i.d, j.o, c.f, n.f, h.l, i.g, i.q, p.g, j.k, d.m, j.p}
@attribute componentW2 {k.f, p.i, f.l, f.p, d.o, r.m, g.f, i.c, s.d, c.n, c.l, p.h, q.n, m.r, m.g, f.e, h.o, g.h,
j.l, o.h, j.m, b.e, h.j, g.m, k.m, h.n, p.n, d.e, p.r, e.s, l.n, j.h, f.k, h.s, h.i, m.i, m.l, l.e, i.o, k.l, r.k, o.f, g.c,
k.b, s.m, n.g, g.j, m.s, h.g, f.h, p.m, d.k, h.a, i.d, j.o, c.f, p.l, n.f, h.l, i.g, i.q, p.g, j.k, d.m, j.p}
@attribute componentW3 {k.f, p.i, f.l, f.p, d.o, r.m, g.f, i.c, s.d, c.n, c.l, p.h, q.n, m.r, m.g, f.e, h.o, g.h,
j.l, o.h, j.m, b.e, h.j, g.m, k.m, h.n, p.n, d.e, p.r, e.s, l.n, f.k, h.s, h.i, m.i, m.l, l.e, i.o, r.k, o.f, g.c, k.b,
s.m, n.g, g.j, m.s, f.h, p.m, d.k, h.a, j.o, c.f, p.l, n.f, h.l, i.g, p.g, j.k, d.m}
@attribute componentW4 {k.f, p.i, f.l, f.p, d.o, r.m, g.f, i.c, s.d, c.n, c.l, p.h, q.n, m.r, m.g, f.e, h.o, g.h,
j.l, o.h, j.m, b.e, h.j, g.m, k.m, h.n, p.n, d.e, p.r, e.s, l.n, j.h, f.k, h.s, h.i, m.i, m.l, l.e, i.o, k.l, r.k, o.f, g.c,
k.b, s.m, n.g, g.j, m.s, h.g, f.h, p.m, d.k, h.a, i.d, j.o, c.f, p.l, n.f, h.l, i.g, i.q, p.g, j.k, d.m}
@attribute Peer {P10, P11, P12, P13, ………………………………., P308, P309}
@data
SP5, Q10, p.r, r.m, m.i, h.i, P114
SP5, Q10, p.r, r.m, m.i, h.i, P263
SP2, Q11, d.e, h.j, m.l, k.m, P39
SP2, Q11, d.e, h.j, m.l, k.m, P253
SP2, Q11, d.e, h.j, m.l, k.m, P91
SP9, Q12, e.s, p.r, g.m, p.n, P247
SP9, Q12, e.s, p.r, g.m, p.n, P130
SP6, Q14, r.k, c.n, d.o, c.l, P87
SP6, Q14, r.k, c.n, d.o, c.l, P117
SP1, Q15, f.p, i.d, b.e, j.o, P29
SP1, Q15, f.p, i.d, b.e, j.o, P44
SP1, Q15, f.p, i.d, b.e, j.o, P56
SP8, Q16, o.f, g.c, l.n, k.b, P213
SP8, Q309, j.h, k.b, l.n, c.f, P279
SP8, Q309, j.h, k.b, l.n, c.f, P209
SP8, Q309, j.h, k.b, l.n, c.f, P149
SP8, Q309, j.h, k.b, l.n, c.f, P16
SP8, Q309, j.h, k.b, l.n, c.f, P168
SP0, Q17, p.i, g.h, k.f, h.j, P224
SP0, Q17, p.i, g.h, k.f, h.j, P73
SP5, Q18, m.i, b.e, p.r, h.s, P256
SP5, Q18, m.i, b.e, p.r, h.s, P43
SP4, Q21, h.o, g.h, j.l, m.r, P246
SP4, Q21, h.o, g.h, j.l, m.r, P301
SP3, Q23, g.f, r.k, f.e, h.i, P53
SP3, Q23, g.f, r.k, f.e, h.i, P286
.....
```

Figure 4. Database in ARFF format

Weka's native storage method is ARFF format, so a conversion has been performed to make the examination data available for analysis through Weka. The most important part of the entire data mining process is preparing the input for data mining investigation. The P2P database contains data from more than 500 Peers with 24 Super-Peers, about 8806 instances (Figure 4, after the line @Data) after a simulation in the architecture-Base, data Extraction and filtering to obtain the ARFF format that is input data to be injected in Weka to obtain the decision tree.

Decision trees are often used in classifications and predictions. It is a simple and powerful way of knowledge representation. The models produced by decision trees are represented in the form of tree structures. A component of query indicates the class of the examples. The instances are classified by sorting them down the tree from the first component of the query to other component of the query.

```
componentW1 = k.f
|   componentW2 = g.f: SP3 (26.0)
|   componentW2 = g.h: SP0 (15.0)
componentW1 = p.i: SP0 (50.0)
componentW1 = f.l: SP1 (78.0/12.0)
componentW1 = f.p: SP1 (159.0/14.0)
componentW1 = d.o
|   componentW4 = r.m: SP3 (38.0/16.0)
|   componentW4 = i.c: SP3 (25.0)
|   componentW4 = s.d: SP6 (28.0)
|   componentW4 = c.l: SP6 (60.0)
|   componentW4 = h.i: SP5 (60.0/37.0)
componentW1 = r.m: SP5 (393.0/138.0)
componentW1 = g.f: SP3 (46.0)
componentW1 = i.c: SP3 (157.0/37.0)
componentW1 = s.d
|   componentW2 = c.n: SP6 (34.0)
|   componentW2 = q.n: SP8 (37.0)
|   componentW2 = f.k: SP6 (31.0)
|   componentW2 = h.l: SP8 (21.0)
|   componentW2 = p.g: SP1 (19.0)
componentW1 = c.n: SP6 (96.0)
componentW1 = q.n
|   componentW2 = m.r: SP7 (140.0)
|   componentW2 = m.g: SP7 (105.0)
|   componentW2 = f.e: SP1 (60.0)
|   componentW2 = j.m: SP1 (26.0)
```

Figure 5. Results of running J48 Decision Tree algorithm

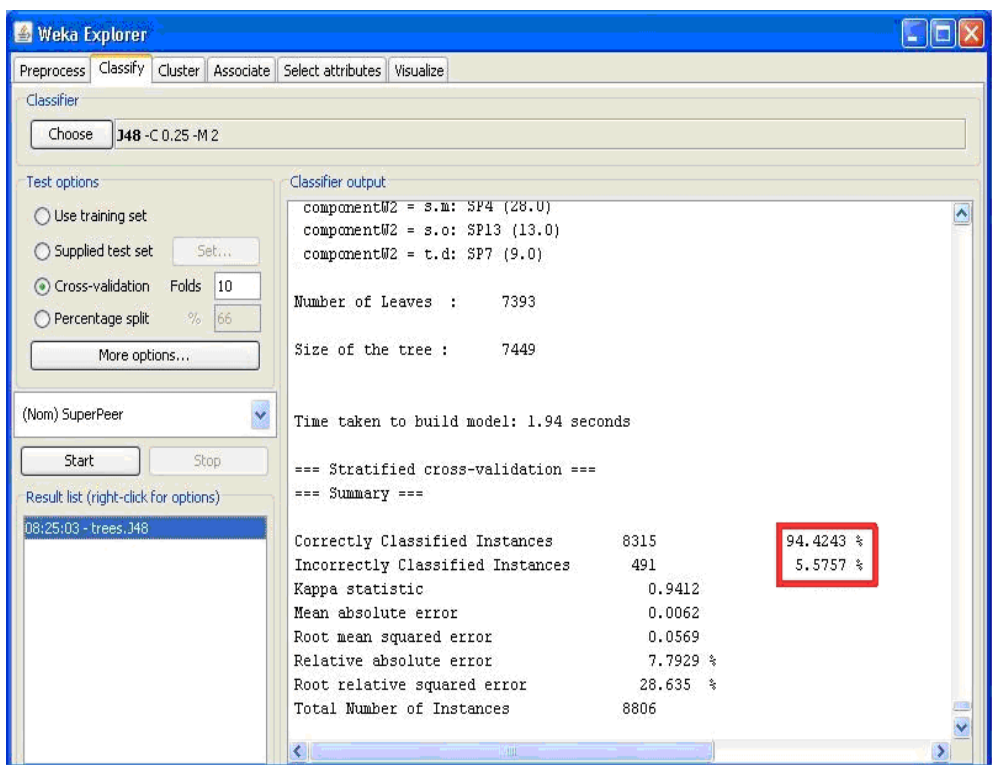

Figure 6. Results of classified instances

Decision trees represent a supervised approach of classifications. Weka uses the J48 algorithm, which is Weka's implementation of C4.5 [32] Decision tree algorithm. J48 is actually a slightly improved version of C4.5. It was the last public version of this family of algorithms before the commercial implementation C5.0 had been released. J48 is the decision tree classification algorithm. It builds a decision tree model by analyzing training data, and uses this model to classify user data. Figure 4 shows the results of running J48 Decision tree algorithm.

The output shown in Figure 5, needs some explanation to see how the tree structure is represented. Each line represents a node in the tree. The lines that start with a 'I', are child nodes of the first line. A node with one or more 'I' character before the rule, is the child node of the node the right most line of 'I' character terminates at. If the rule is followed by a colon and a class designation, then that designation becomes the classification of the rule. If it isn't followed by a colon, continue to the next node in the tree.

Classification of large datasets is an important data mining methodology. For our purposes, the most important figures here are the numbers of correctly and incorrectly classified instances. The output from the Weka program is shown in Figure 6. In this output, the decision tree is able to correctly classify approximately ninety two percent of the data.

## 6. RESULTS AND DISCUSSION

Evaluating the performance of P2P network is an important part in understanding how useful it can be in the real world. As with all P2P applications, the first question is whether P2P is scalable or it is not. Our system was evaluated with different set of parameters i.e. number of Peers, number of Super-Peer etc. Evaluation results were quite encouraging. There are many dimensions in which scalability can be evaluated: one important metric is the time that takes the Answer to a given query. We ran simulations on P2P network in three different sizes. Each Peer sends Query to his Super-Peer that sends the query to a Super-Super-Peer in order to precisely identify which Super-Peer(s) can answer the given query.

- In the first run, we fixed the number of Super-Peers (10 Super-Peers) and we modified the number of Peers (500, 1000, 1500, ….., 3000 Peers) in both Architectures to measure the execution time.

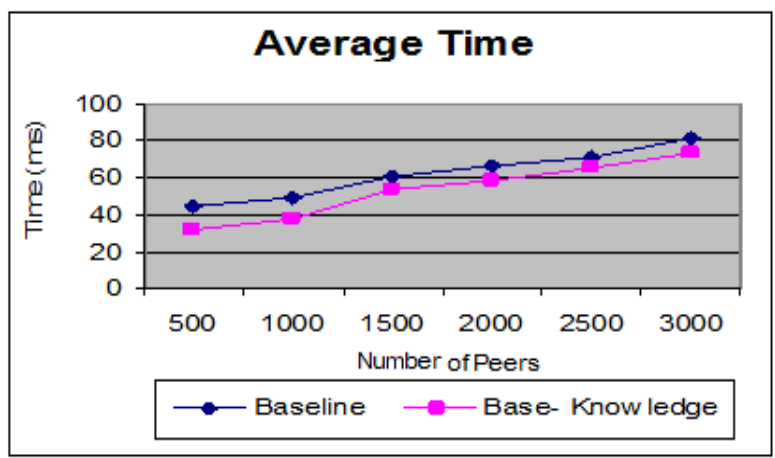

Figure 7. Evolution time in both architectures when increasing the Peers

- In the second one, we fixed the number of Peers (3000 Peers) and we modified the number of the Super-Peers (4, 5, 6,…., 24) in both architectures to measure the execution time.

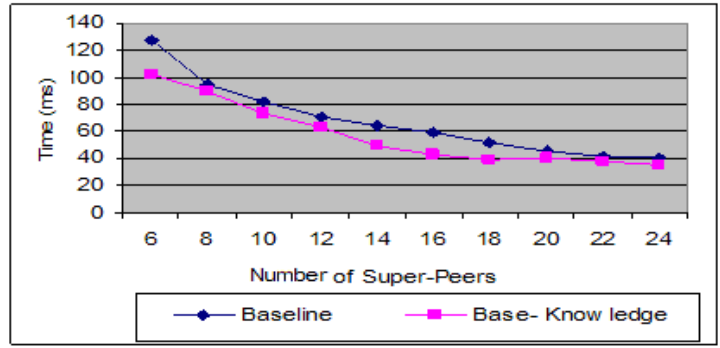

Figure 8. Evolution time in both architectures when increasing the Super-Peer

- In the third one, we modified the number of Peers (500, 1000, 1500, ….., 3000 Peers) and Super-Peers (4,5,6,…., 24) in both architectures to measure the execution time.

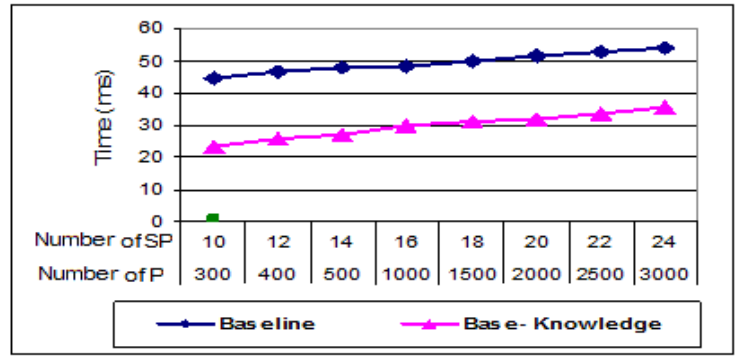

Figure 9. Evolution time in both architecture when increasing the Super-Peers and Peers

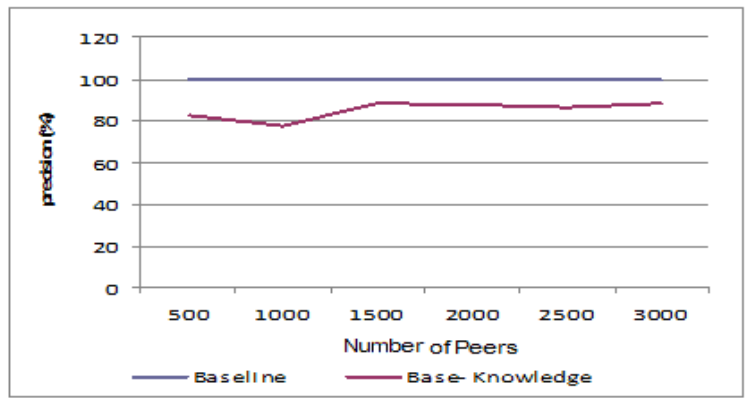

Figure 10. Precision between both architectures

The graphs shown in figures 7, 8, 9 and 10 are the results of our simulations. They demonstrate the performance of using the Super-Super-Peer with Decision Tree for the discovery of P2P Domain (SP). The considered time is affected by the increase of the increase of the repository size. Measurements depicted in Figures 7 and 9, show that the publishing time has increased in architecture base-Knowledge is less than that in the Baseline architecture. Figure 8 shows that the time decreased in architecture base-Knowledge is more than that of the baseline architecture when the number of Super-Peers is increased.

Measurements in Figure 10 show the precision (%) of the architecture base-Knowledge as compared to architecture-Base, which is our baseline.

Precision (in %) (For each Query) = [(Number of Peers obtained for such Query)/(Number of Peers that answered in the baseline architecture)] * 100

We could observe that there is almost a linear line in precision (number of Peers more then 2000) between architectures, which reflects the stability of our architecture while increasing the number of Peers. This experiment was designed to measure the accuracy of data (since precision is almost not affected by the network size).

Finally, our prototype in predicting P2P domain raises some interesting performance issues while using decision tree. We have performed experiments to demonstrate how the presence of multiple domains affects the performance. These experiments also illustrate how our method can improve the scalability of the overall system (number of Peers and Super-Peers increased at the same time). These results showed that our proposed processing strategy is efficient for Super-Peer based network.

## 7. CONCLUSIONS & FUTURE WORK

Classification based on decision trees is one of the important problems in data mining. In recent years, database systems have become highly distributed. The distributed system paradigms, such as Peer-to-Peer databases, are being adopted. Initially, the experiments have been conducted on the whole P2P dataset. The reason for selecting the J48 decision tree algorithm is because the algorithm has the ability to handle data with missing attribute values better than ID3 decision tree algorithm. It also avoids over fitting the data and reduces error pruning. The experiments involved more than 3000 Peers with 24 Super-Peers. The decision tree is useful to solve a particular problem, and to form the basis of evaluating the performance of Queries Answering in P2P network. The advantage of this model is the robustness in Queries Answering and scalability issues in P2P Network, at the same time respecting very importing issues such as data privacy and the dynamic nature of the underlying network: Peers can leave the overlay network and new Peers can join it.

One important area for improvement is performance. Some of the options for improving performance were discussed in the evaluation of P2P Network and include: improvements in the answering time, a given query and dynamic nature of P2P Network.

In this paper, we investigated P2P systems that are currently in use, primarily on decentralized and unstructured systems. The unstructured systems are actively used by the largest community of Internet users and support many desirable properties. Two major deficiencies of unstructured P2P networks are addressed: scalability and efficient search mechanisms.

Consequent to our observations, we propose a hierarchical-based Super-Peer structure. Super-Peers are then selected from regular Peers to act as cluster leaders, responsible for locating content and maintaining the network structure for client Peers.

Super-Peers are also connected to each other, forming a Super-Peer overlay network. To scale the routing on the overlay network connecting the Super-Peer nodes that are connected to Super-Super-Peer according to their semantic domains, we have constructed a predictive model that avoids flooding queries in the P2P network by predicting the appropriate Super-Peer, and hence the Peer to answer the query.

Another major direction for future work is enhancing the performance (Answering time) by grouping the Domains (Super-Peers) into multiples Super-Super-Peers to minimize the load on one Super-Super-Peer and also to be more scalable.

## Authors

**Dr. Anis Ismail**, Born in Lebanon, works as system and network administrator and instructor at the Lebanese University, University Institute of technology , Sidon, Lebanon. He has a B.S. degree in Telecommunication and Networking Engineering from the Lebanese University (LU), an M.S. in Computer Science from the American University of Science and Technology (AUST) in Lebanon, and a Ph.D. in Computer Science from the University of AIX-Marseille, France. His main research interest covers Data Mining in P2P Systems, Arabic Language Processing, and Multimedia Information.

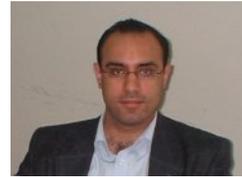

**Dr. Aziz M. Barbar** is the Chairperson of the Department of Computer Science at the American University of Science & Technology (AUST), Lebanon. He has a Ph.D. in Computer Science from the University of Nice-Sophia Antipolis (France). His researc h interests include Database Reverse Engineering, Data Mining and Natural Language Processing. Dr. Barbar is currently the Vice-President of the Lebanese Information Technology Association (LITA), and the Chair of the IEEE Computer Chapter, Lebanon.

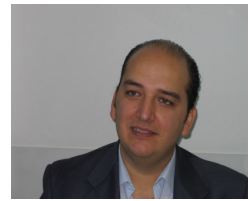

**Ziad Imail,** Born in Lebanon, has a Bachelor and a Master of Engineering in Telecommunicati-ons and Networking from respectively the University of Saint-Joseph and Telecom Paristech. His research interests cover Networking and System Security.

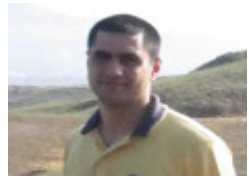